\documentclass{moriond}

\usepackage{graphicx}
\usepackage{hyperref}
\usepackage{amsmath, amssymb}
\usepackage{babel}
\usepackage{color}
\usepackage{mathrsfs}
\usepackage{slashed}

\def\be{\begin{equation}}
\def\ee{\end{equation}}
\def\bea{\begin{eqnarray}}
\def\eea{\end{eqnarray}}

\newcommand{\Photo}{\includegraphics[height=30mm]{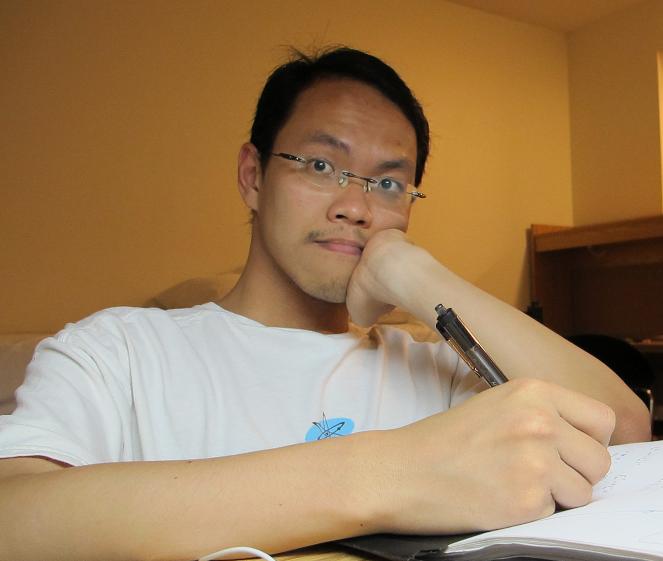}}

\begin{document}
\vspace*{4cm}
\title{NEW CALCULATIONS OF NEUTRON AND KAON DECAYS}

\author{CHIEN-YEAH SENG}
	
\address{Helmholtz-Institut f\"{u}r Strahlen- und Kernphysik and Bethe Center for
		Theoretical Physics,\\ Universit\"{a}t Bonn, 53115 Bonn, Germany}
	
\maketitle\abstracts{We describe two recent developments of the radiative correction theory in semileptonic beta decays. For neutron, a novel dispersive analysis of the $\gamma W$-box diagram making use of neutrino scattering data leads to an improved precision and a shifted central value of $V_{ud}$; for kaon, an appropriate combination of meson form factors, lattice QCD inputs and chiral perturbation theory gives a much-improved determination of the radiative corrections in semileptonic kaon decays. These new calculations help to unveil/sharpen a series of anomalies in the measurements of $V_{ud}$ and $V_{us}$ that may point towards the existence of new physics. }
	
\section{Introduction}

\begin{figure}
	\begin{centering}
		\includegraphics[scale=0.35]{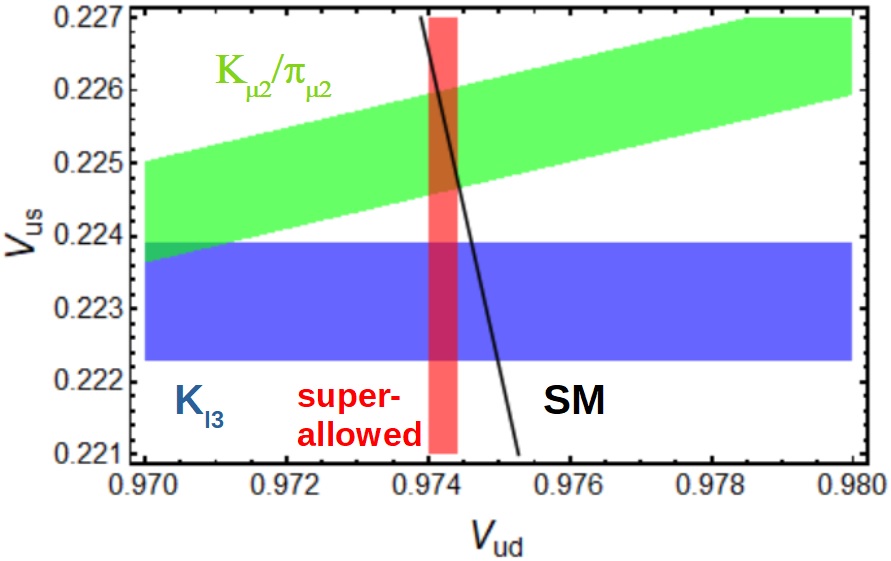}
		\includegraphics[scale=0.35]{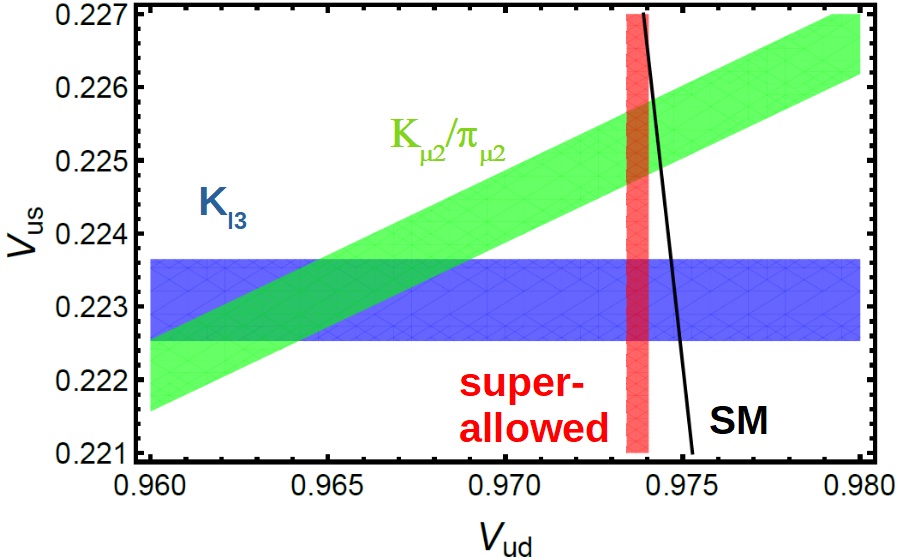}
		\hfill
		\par\end{centering}
	\caption{\label{fig:VudVus}Summary of $V_{ud}$ and $V_{us}$ determined from superallowed beta decays, $K_{\ell 3}$ and $K_{\mu 2}/\pi_{\mu 2}$ in early 2018 (left) and now (right). The black line represents the SM prediction of the first-row CKM unitarity. }
\end{figure}

Leptonic and semileptonic beta decay processes are primary avenues to extract the first-row Cabibbo-Kobayashi-Maskawa (CKM) matrix elements: $V_{ud}$ is best measured from superallowed $0^+\rightarrow 0^+$ nuclear decays, $V_{us}$ is best measured from semileptonic kaon decays ($K_{\ell 3}$), while the ratio $V_{us}/V_{ud}$ is best measured from the $K\rightarrow\mu\nu$ ($K_{\mu 2}$) and $\pi\rightarrow\mu\nu$ ($\pi_{\mu 2}$) leptonic decays. The left panel in Fig.\ref{fig:VudVus} summarizes the values of $V_{ud}$ and $V_{us}$ extracted from these processes until early 2018~\cite{ParticleDataGroup:2018ovx}. These measurements allow us to check some of the Standard Model (SM) predictions, for example the first-row CKM unitarity, which reduces to a simple Cabibbo unitarity $|V_{ud}|^2+|V_{us}|^2=1$ upon neglecting the very small $|V_{ub}|^2$. Notice, however, that a good control of all the SM theory inputs to these decay processes is required for high-precision extractions of the CKM parameters; and a class of the most important SM corrections is the electroweak radiative corrections (RC) which involves both virtual loop corrections and real (bremsstrahlung) corrections. The non-perturbative strong interaction of the decayed particles contributes to a large hadronic uncertainty in the RC of these decay processes.

\begin{figure}
	\begin{centering}
		\includegraphics[scale=0.45]{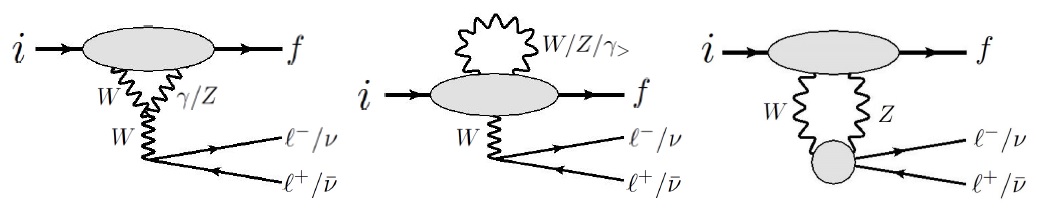}
		\includegraphics[scale=0.45]{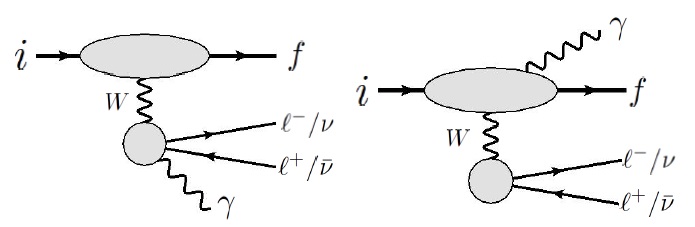}
		\includegraphics[scale=0.45]{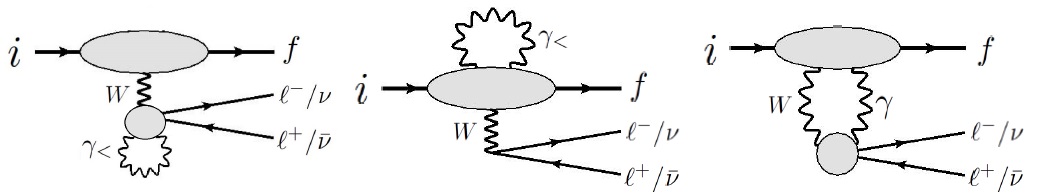}
		\hfill
		\par\end{centering}
	\caption{\label{fig:EWRC}Feynman diagrams representing the weak RC (the first three diagrams), bremsstrahlung (the fourth and fifth diagrams) and the long-distance virtual electromagnetic RC (the last three diagrams) in a generic semileptonic beta decay. Here, $\gamma=\gamma_<+\gamma_>$ denotes the splitting of the full photon propagator into ``massive'' and ``massless'' pieces, see Ref.~\protect\cite{Sirlin:1977sv}.  }
\end{figure}	

Recent years have seen a significant development in the understanding of these RC based on a resurrected, classical theory framework known as ``Sirlin's representation'' which was originally formulated by Alberto Sirlin in late 1970s~\cite{Sirlin:1977sv} to study the electroweak RC in superallowed beta decays; the recent development extends the method to RC in generic semileptonic beta decays~\cite{Seng:2021syx}. Within this framework, the full $\mathcal{O}(G_F\alpha)$ RC is divided into three parts (see Fig.\ref{fig:EWRC}):
\begin{itemize}
	\item The ``weak'' RC which probes only physics at the ultraviolet (UV) scale ($q\sim M_W$, where $q$ is the loop momentum) and are calculable to satisfactory precision using perturbative Quantum Chromodynamics (pQCD). 
	\item The bremsstrahlung correction involves the emission of a real photon. It probes only the low-energy electromagnetic (EM) interactions that are well-constrained by experimental data.
	\item The long-distance virtual EM correction involves photon loop diagrams. This is the part that contains the large hadronic or nuclear uncertainties. 
\end{itemize}
Current algebra allows one to further subdivide the non-trivial virtual EMRC into two classes: The first class depends on physics at $Q^2=-q^2\ll 1$~GeV$^2$ which again can be fixed by a small amount of inputs; the second class, which comes from the so-called $\gamma W$-box diagram (the last diagram in Fig.\ref{fig:EWRC}), is sensitive to physics at all scales including $Q^2\sim 1$~GeV$^2$, i.e. the hadronic scale, where QCD becomes non-perturbative. It thus carries a large hadronic uncertainty and requires special treatments using extra computational techniques; once this piece is under control, a precision level of the order $10^{-4}$ for the RC to the interested decay process can be achieved.

\section{RC in free neutron decay}

\begin{figure}
	\begin{centering}
		\includegraphics[scale=0.2]{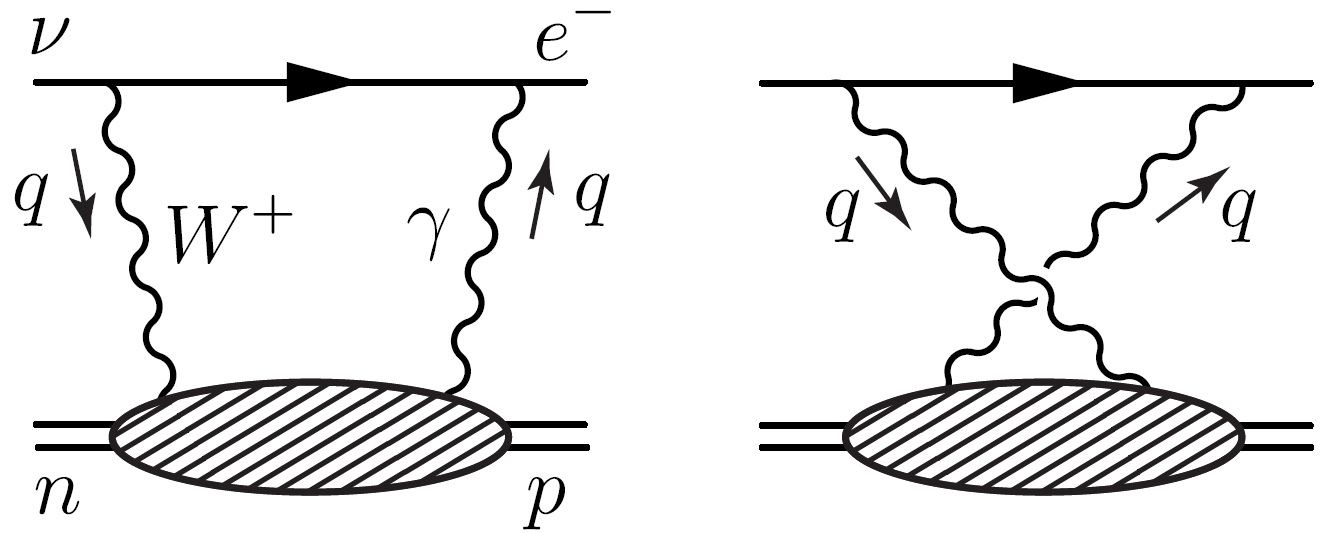}
		\includegraphics[scale=0.14]{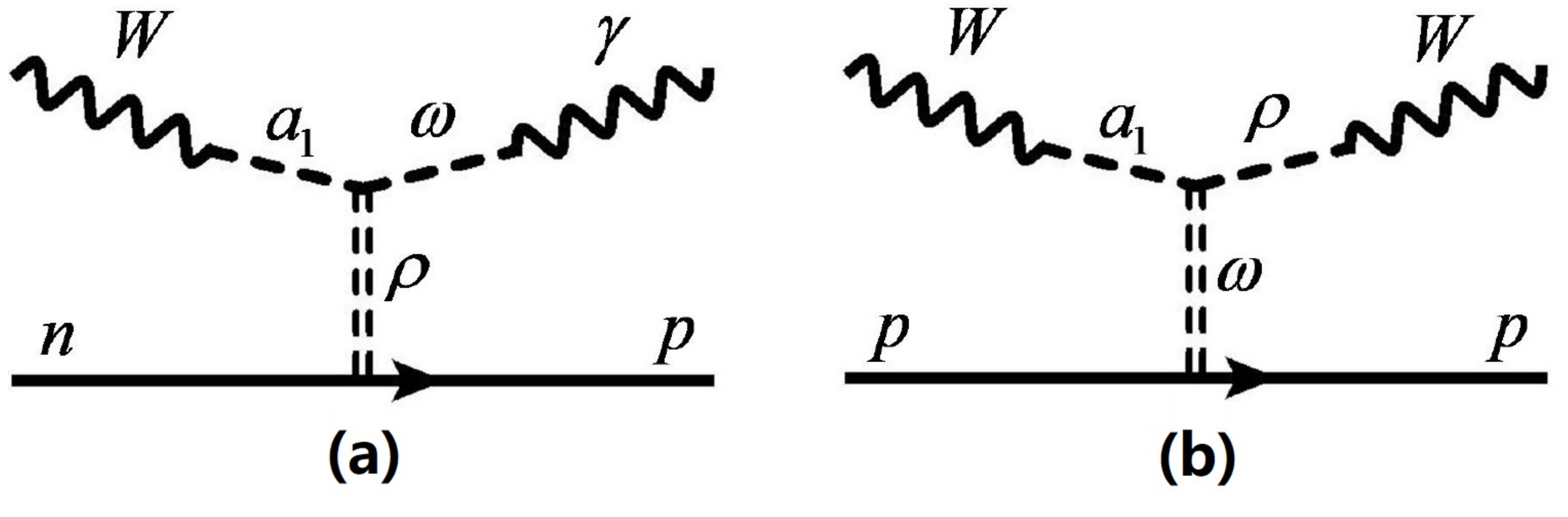}
		\hfill
		\par\end{centering}
	\caption{\label{fig:box}First pair: The single-nucleon $\gamma W$-box diagram. Second pair: The leading Regge-exchange picture for the $nW\rightarrow p\gamma$ and $pW\rightarrow pW$ processes respectively.  }
\end{figure}

Sirlin's representation is particularly advantageous in near-degenerate decays, i.e. the initial and final strongly-interacting particles are almost degenerate; examples are semileptonic decays of pion, free neutron and nuclear systems. In this limit, many terms in Sirlin's representation either vanish or are saturated by the exactly-calculable ``convection term contribution'' (which gives rise to the so-called ``outer corrections''), where the photon couples only to the hadrons' electric charge. Meanwhile, the forward limit ($p_f\rightarrow p_i$) can be taken for the remaining terms which significantly simplifies the integral structure. 

An important example of such kind is the free neutron beta decay which plays a central role in the $V_{ud}$ extraction (either by itself or as an integrated component of nuclear beta decays). Sirlin's representation states that the primary source of theory uncertainty in this decay is the RC originates from the ``single-nucleon forward axial $\gamma W$-box diagram'' depicted by the first pair of diagrams in Fig.\ref{fig:box}, where the W-boson couples to the non-conserved axial current at the nucleon side. The incalculable hadron physics probed by the loop integral at $Q\sim 1$~GeV constitutes a large theory uncertainty in $V_{ud}$. 

In year 2018, a novel dispersion relation (DR) treatment~\cite{Seng:2018yzq,Seng:2018qru} was introduced to relate the integrand of the box diagram to experimental observables. For the correction to the free neutron vector coupling $g_V$ relevant to the $V_{ud}$ extraction, DR gives:
\begin{equation}
\Box_{\gamma W}^V=\frac{\alpha}{\pi g_V}\int_0^\infty\frac{dQ^2}{Q^2}\frac{M_W^2}{M_W^2+Q^2}\int_0^1 dx\frac{1+2r}{(1+r)^2}F_3^{(0)}(x,Q^2)~,
\end{equation}
where $r=\sqrt{1+4m_N^2x^2/Q^2}$, and $F_3^{(0)}(x,Q^2)$ is the parity-odd structure function originates from the interference between the axial charged weak current and the isoscalar (hence the superscript $(0)$) component of the EM current. It is not directly measurable, but making use of a Regge-exchange picture (the second pair of diagrams in Fig.\ref{fig:box}), the contribution from the non-elastic intermediate states to $\Box_{\gamma W}^V$ can be largely fixed by the structure function $F_3^{\nu p+\bar{\nu}p}$ measured from neutrino/antineutrino-nucleus scattering. Making use of currently-available data, the new analysis reduced the uncertainty in the previous state-of-the-art determination of $\Box_{\gamma W}^V$~\cite{Marciano:2005ec} by a factor 2, but also observed a significant increase of its central value which shifted the value of $|V_{ud}|$ from 0.97420(21) in early 2018 to 0.97370(14) in late 2018. This new discovery unveiled, for the first time in the past few years, the tension in the first-row CKM unitarity. This observation was later confirmed in several independent studies~\cite{Czarnecki:2019mwq,Seng:2020wjq,Shiells:2020fqp,Hayen:2020cxh}. Nevertheless, later studies also discovered new sources of theory uncertainties from nuclear structure effects~\cite{Seng:2018qru,Gorchtein:2018fxl} which lead to the current value of $|V_{ud}|=0.97373(31)$ from superallowed decays~\cite{Hardy:2020qwl}.

The precision of the DR analysis is limited by the available experimental data, and first-principles calculations of the single-nucleon box diagram are needed to overcome this natural limitation. A first prototype was made in Ref.~\cite{Feng:2020zdc} where a simpler box diagram of charged pions was computed to percent-level accuracy by combining pQCD results at large-$Q^2$ and lattice calculations of four-point correlation functions at small-$Q^2$. First results for the single-nucleon box diagram using similar techniques or alternative approaches~\cite{Seng:2019plg} may be expected within the next few years.

\section{RC in kaon semileptonic decays} 

\begin{figure}
	\begin{centering}
		\includegraphics[scale=0.45]{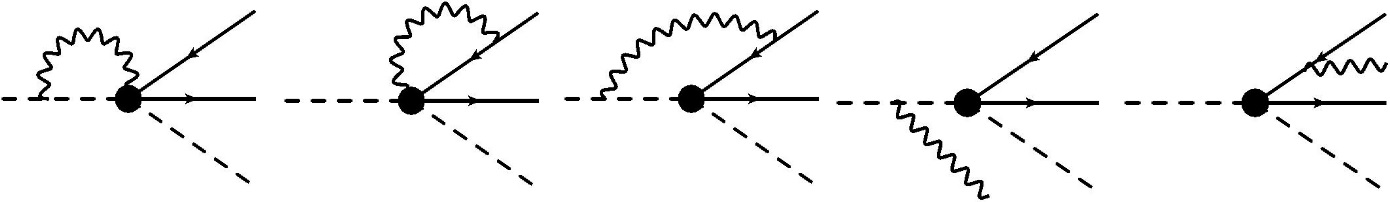}
		\hfill
		\par\end{centering}
	\caption{\label{fig:ChPT}Feynman diagrams of the $K_{\ell 3}$ RC computed with ChPT. }
\end{figure}	

With its success in neutron, Sirlin's representation was later applied to study the RC in $K_{\ell 3}$ decays, i.e. $K\rightarrow\pi\ell^+\nu(\gamma)$, which are crucial in the extraction of $V_{us}$~\cite{Seng:2019lxf,Seng:2020jtz,Ma:2021azh,Seng:2021boy,Seng:2021wcf,Seng:2022wcw}.
It is more complicated than the free neutron RC due to several reasons: (1) Unlike neutron, many terms in Sirlin's representation either do not vanish, or are not saturated by the convection terms contribution; and (2) the forward limit cannot be simply taken because the kaon and pion momenta are very different. Previous state-of-the-art computation of this RC was based on Chiral Perturbation Theory (ChPT)~\cite{Cirigliano:2008wn}, which was done by computing the virtual and real emission diagrams in Fig.\ref{fig:ChPT} using an effective Lagrangian. 
Two major sources of theory uncertainties within this framework are: (1) Uncertainties due to neglected higher-order terms ($\mathcal{O}(e^2p^{2n})$, where $n\geq 2$) in chiral power counting, and (2) uncertainties due to poorly-known low energy constants (LECs). They both contribute to a theory error of the order $10^{-3}$.

To make full use of Sirlin's representation, two important observations are in order. First, the tree-level decay amplitude is completely fixed by two $K\pi$ charged weak form factors $f_\pm$:
\begin{equation}
\langle \pi(p')|(J_\mu^W)^\dagger|K(p)\rangle=V_{us}^*[f_+(t)(p+p')_\mu+f_-(t)(p-p')_\mu]~,
\end{equation} 	
(where $t=(p-p')^2$) 
and the contribution of $f_-(t)$ to the decay rate is suppressed by $m_\ell^2/M_K^2$. Second, the virtual corrections can always be expressed as corrections to the form factors, i.e.,
\begin{equation}
f_\pm(t)\rightarrow f_\pm(t)+\delta f_\pm(y,z)~,
\end{equation}
(where $y=2p\cdot p_\ell/M_K^2$ and $z=2p\cdot p'/M_K^2$) and the contribution of $\delta f_-$ to the decay rate is suppressed by the same kinematic factor. This leads us to the following simple strategy: On the one hand, one may calculate $\delta f_+$ with fully non-perturbative methods based on Sirlin's representation; on the other hand, for $\delta f_-$ which is only relevant to $K_{\mu 3}$, one may split it into infrared (IR)-singular and IR-regular pieces, the former is calculated non-perturbatively and the latter perturbatively. Based on this strategy, a re-analysis was performed first to the RC in $K_{e3}$ decays~\cite{Seng:2021boy,Seng:2021wcf} and later in $K_{\mu 3}$~\cite{Seng:2022wcw}, which we briefly describe as follows. We start from $\delta f_+$ which are computed non-perturbatively:
\begin{enumerate}
\item The first non-trivial contribution to $\delta f_+$ comes from following loop integral:
\begin{eqnarray}
I_\mathfrak{A}^\lambda&=&-e^2\int\frac{d^4q'}{(2\pi)^4}\frac{1}{[(p_\ell-q')^2-m_\ell^2][q^{\prime 2}-M_\gamma^2]}\left\{\frac{2p_\ell\cdot q'q^{\prime\lambda}}{q^{\prime 2}-M_\gamma^2}T^\mu_{\:\:\mu}+2p_{\ell\mu}T^{\mu\lambda}\right.\nonumber\\
&&\left.-(p-p')_\mu T^{\lambda\mu}+i\Gamma^\lambda-i\epsilon^{\mu\nu\alpha\lambda}q_\alpha'(T_{\mu\nu})_V\right\}~,
\end{eqnarray}
where $T_{\mu\nu}$ and $\Gamma_\lambda$ are tensors involving time-ordered product of the EM and charged-weak current, and the subscript $V$ denotes the vector charged weak current contribution.
It turns out that this integral is insensitive to UV physics and is practically saturated by the ``pole term'' contribution to $T_{\mu\nu}$ and $\Gamma_\lambda$ which requires only the knowledge of the $K/\pi$ EM and charged weak form factors. Substituting the full form factors into the integral is equivalent to re-summing the most important $\mathcal{O}(e^2p^{2n})$ corrections in ChPT. 
\item Next is the axial current contribution to the (non-forward) $\gamma W$-box diagram, which is contained in the following integral:
\begin{equation}
I_\mathfrak{B}^\lambda=ie^2\int\frac{d^4q'}{(2\pi)^4}\frac{M_W^2}{M_W^2-q^{\prime 2}}\frac{\epsilon^{\mu\nu\alpha\lambda}q'_\alpha(T_{\mu\nu})_A}{[(p_\ell-q')^2-m_\ell^2]q^{\prime 2}}~.
\end{equation}
It may be pinned down by computing an unphysical, forward $K\rightarrow\pi$ axial $\gamma W$-box diagram in the degenerate limit, $M_K=M_\pi$. The full box diagram splits into two pieces:
\begin{equation}
\Box_{\gamma W}(K,\pi,M_\pi)=\Box_{\gamma W}^>+\Box_{\gamma W}^<(K,\pi,M_\pi)~,
\end{equation}
where the ``$>$'' component is calculated from pQCD, while the ``$<$'' component is calculated directly with lattice QCD~\cite{Ma:2021azh,Seng:2021qdx}. With that one may write:
\begin{equation}
(\delta f_+)_\mathfrak{B}=\Box_{\gamma W}^> f_+(t)+\{\Box_{\gamma W}^<(K,\pi,M_\pi)+\mathcal{O}(M_K^2/\Lambda_\chi^2)\}f_+(t)~.
\end{equation}
The theory uncertainty due to non-forward corrections is only associated to $\Box_{\gamma W}^<$, and could be estimated using the standard chiral power counting argument. 
\item Next we have the so-called ``three-point function'' contribution to $\delta f_+$ which we will not rigorously define here; what we need to know is that it splits into two pieces:
\begin{equation}
\delta f_{+,3}=(\delta f_+)_\mathrm{III}+(\delta f_{+,3})^\mathrm{fin}~,
\end{equation}	
where the first term is IR-divergent but exactly calculable, while the second term is IR-finite and may adopt a fixed-order ($\mathcal{O}(e^2p^2)$) ChPT approximation. Moreover, it turns out that the latter is usually not included in the definition of long-distance EM corrections according to the standard ChPT treatment, but is rather reabsorbed into tree-level form factors and isospin-breaking corrections.
\end{enumerate}

This completes the study of $\delta f_+$. For $\delta f_-$, we again split it into different components:
\begin{equation}
\delta f_-=(\delta f_-)_\mathrm{I+II+III}+(\delta f_-)_\mathrm{conv}^\mathrm{fin}+(\delta f_-)_\mathrm{rem}~,
\end{equation}	
where I+II+III and ``conv(fin)'' are IR-divergent, numerically large but exactly calculable in Sirlin's representation; the remaining (``rem'') component is incalculable within this framework but its contribution to the decay rate is numerically small. Since the $\mathcal{O}(e^2p^2)$ expression of the full $\delta f_-$ is available in the literature~\cite{Cirigliano:2001mk}, one could also deduce the $\mathcal{O}(e^2p^2)$ approximation for $(\delta f_-)_\mathrm{rem}$, which is sufficient for our precision goal. 
Finally we have the bremsstrahlung contribution which is again separated into two parts:
\begin{equation}
\mathfrak{M}_\mathrm{brems}=\mathfrak{M}_A+\mathfrak{M}_B~,
\end{equation}
where $\mathfrak{M}_A$ contains the full IR-divergent structure due to the convection term contribution and is exactly calculable, while $\mathfrak{M}_B$ is IR-finite and may adopt a fixed-order ChPT approximation.

\begin{table}
	\begin{centering}
		\begin{tabular}{|c|c|c|}
			\hline 
			& $\delta_{\mathrm{EM}}^{K\ell}$ & ChPT\tabularnewline
			\hline 
			\hline 
			$K^{0}e$ & $11.6(2)_{\mathrm{inel}}(1)_{\mathrm{lat}}(1)_{\mathrm{NF}}(2)_{e^{2}p^{4}}$ & $9.9(1.9)_{e^{2}p^{4}}(1.1)_{\mathrm{LEC}}$\tabularnewline
			\hline 
			$K^{+}e$ & $2.1(2)_{\mathrm{inel}}(1)_{\mathrm{lat}}(4)_{\mathrm{NF}}(1)_{e^{2}p^{4}}$ & $1.0(1.9)_{e^{2}p^{4}}(1.6)_{\mathrm{LEC}}$\tabularnewline
			\hline 
			$K^{0}\mu$ & $15.4(2)_{\mathrm{inel}}(1)_{\mathrm{lat}}(1)_{\mathrm{NF}}(2)_{\mathrm{LEC}}(2)_{e^{2}p^{4}}$ & $14.0(1.9)_{e^{2}p^{4}}(1.1)_{\mathrm{LEC}}$\tabularnewline
			\hline 
			$K^{+}\mu$ & $0.5(2)_{\mathrm{inel}}(1)_{\mathrm{lat}}(4)_{\mathrm{NF}}(2)_{\mathrm{LEC}}(2)_{e^{2}p^{4}}$ & $0.2(1.9)_{e^{2}p^{4}}(1.6)_{\mathrm{LEC}}$\tabularnewline
			\hline 
			\end{tabular}
			\par\end{centering}
			\caption{\label{tab:Kl3final}Comparison between the long-distance EMRC determined by Sirlin's representation
				and pure ChPT.}
				
				\end{table}

With the strategies above, Refs.~\cite{Seng:2021boy,Seng:2021wcf,Seng:2022wcw} updated the theory predictions of long-distance EMRC to all channels in $K_{\ell 3}$ decays, see Table~\ref{tab:Kl3final}. The main sources of theory uncertainty in the new determination are: (1) Contributions from inelastic (inel) states to $I_\mathfrak{A}^\lambda$, (2) Lattice (lat) uncertainty in $\Box_{\gamma W}^<$, (3) Uncertainty due to non-forward (NF) effects in the $\gamma W$-box diagram, (4) Higher-order ($\mathcal{O}(e^2p^4)$) ChPT corrections and, (5) The poorly-determined LECs in $\delta f_-$. The new results agree with the pure ChPT results within error bars, but have a much higher theory precision at the level of $10^{-4}$ in contrast to $10^{-3}$ from ChPT.

\section{Conclusion}
	
We have discussed several recent improvements in the study of the RCs in neutron and semileptonic kaon decays. Sirlin's representation provides a useful framework to separate pieces with large theory uncertainties from other parts that are precisely calculable; extra computational techniques are then applied to reduce the theory errors in the former. For neutron, the forward axial $\gamma W$-box diagram was studied with DR techniques which make use of neutrino scattering data to pin down the contributions from inelastic intermediate states. For $K_{\ell 3}$, the full RC is split into terms that are unsuppressed/suppressed in the decay rate and are studied separately. The unsuppressed terms are calculated non-perturbatively using elastic form factors and lattice QCD inputs, while the suppressed terms are computed to fixed order in ChPT. 

The new treatments have significantly reduced the RC uncertainties in the respective processes. Combining with other improvements, the new status of $V_{ud}$ and $V_{us}$ is given by the right plot in Fig.\ref{fig:VudVus}. Unlike that in year 2018, we now observe a number of discrepancies between different extractions of these parameters, as well as systematic deviations from the SM prediction of first-row unitarity. These anomalies are generically at the level of 3$\sigma$, and future progress from both the theory and experiment are required to understand whether they represent signals of physics beyond the SM, or rather some unexpectedly large SM corrections.

\section*{Acknowledgments}
	
The author thanks the organizers of the ``56$^\mathrm{th}$ Rencontres de Moriond'' for the kind invitation, and is grateful to X. Feng, D. Galviz, M. Gorchtein, L. Jin, P. Ma, W. Marciano, U.-G. Mei{\ss}ner, H. Patel and M.J. Ramsey-Musolf for collaborations.
The work of the author is supported in
part by the Deutsche Forschungsgemeinschaft (DFG, German Research
Foundation) and the NSFC through the funds provided to the Sino-German Collaborative Research Center TRR110 ``Symmetries and the Emergence of Structure in QCD'' (DFG Project-ID 196253076 - TRR 110, NSFC Grant No. 12070131001).

\end{document}